\documentclass[12pt,a4paper]{article}
\usepackage[cp1251]{inputenc}
\usepackage{epsfig}
\usepackage{amssymb}
\usepackage{amsmath}
\usepackage{color}

\newcommand{\bm}[1]{{\boldsymbol{\bf #1}}}
\renewcommand{\vec}[1]{{\bm{#1}}}
\renewcommand{\Vec}[1]{{\mathcal{#1}}}
\newcommand{\fr}[2]{{\displaystyle \frac{#1}{#2}}}
\newcommand{\sfr}[2]{{{#1}/{#2}}}

\newcommand{\pdiff}[2]{{\fr{\partial{#1}}{\partial{#2}}}}

\newcommand{\spdiff}[2]{{\sfr{\partial{#1}}{\partial{#2}}}}

\newcommand{\ff}{{{f}{\!}{f}}}

\title%
{%
Modeling of Protostellar Clouds and their Observational Properties
}%
\author%
{%
 A.G. Zhilkin$^{1,2}$\thanks{E-mail: zhilkin@inasan.ru}, \ %
 Ya.N. Pavlyuchenkov$^{1}$, \ %
 S.N. Zamozdra$^{2}$\\ %
\textit{\small $^{1}$ Institute of Astronomy, Russian Academy of Sciences, Moscow, Russia}\\%
\textit{\small $^{2}$ Chelyabinsk State University, Chelyabinsk, Russia}\\%
}%
\date{}

\begin{document}

\maketitle%

\begin{abstract}
\noindent A physical model and two-dimensional numerical method for computing the evolution and spectra of protostellar clouds are described. The physical model is based on a system of magneto-gasdynamical equations, including ohmic and ambipolar diffusion, and a scheme for calculating the thermal and ionization structure of a cloud. The dust and gas temperatures are determined during the calculations of the thermal structure of the cloud. The results of computing the dynamical and thermal structure of the cloud are used to model the radiative transfer in continuum and in molecular lines. We presented the results for clouds in hydrostatic and thermal equilibrium. The evolution of a rotating magnetic protostellar cloud starting from a quasi-static state is also considered. Spectral maps for optically thick lines of linear molecules are analyzed. We have shown that the influence of the magnetic field and rotation can lead to a redistribution of angular momentum in the cloud and the formation of a characteristic rotational velocity structure. As a result, the distribution of the velocity centroid of the molecular lines can acquire an hourglass shape. We plan to use the developed program package together with a model for the chemical evolution to interpret and model observed starless and protostellar cores.
\end{abstract}

\section{Introduction}

Cool, dense, quasi-spherical condensations of gas and dust that do not contain infrared sources of radiation can often be distinguished in the observed complex structures of molecular clouds --- so-called starless cores (see, for example \cite{Bergin2007}). These objects are characterized by densities of $10^4$-$10^6$~cm$^{-3}$, temperatures of 8-20~K, sizes of $0.1$~pc, and masses
from fractions of a solar mass to several tens of solar masses. It is believed that, in the absence of disruptive influences, they can enter a state of gravitational collapse that leads to the birth of young stars. In this case, these objects are usually called pre-stellar cores. Pre-stellar cores can evolve into protostellar cores, in which infrared sources are observed. We will use the term \textit{protostellar cloud} as a gravitationally bound gas-dust condensation whose evolution leads to the formation of a single star or multiple star system. In various stages of its evolution, such a protostellar
cloud can be manifest observationally as a starless or protostellar core.

Starless and protostellar cores are observed primarily via the continuum (thermal) emission of dust and lines of compound molecules (CO, HCO$^{+}$,
N$_2$H$^{+}$ and others). Maps of dust emission are used to establish the density distribution, while the molecular line emission provides information about the thermal, chemical, and dynamical structure of the cores (see,
for example, \cite{Evans1999, Bergin2007}). Both analytical (for example,
the self-similar solutions \cite{Larson1969, Shu1977, Dudorov2008}) and numerical (\cite{Larson1969, Black1975, Matsumoto2004}; see also the review \cite{Klein2007}) models have been actively used in theoretical studies of the evolution of protostellar clouds. Numerical modeling has been used to study various effects and processes that influence the evolution of protostellar clouds, such as magnetic fields, turbulence, rotation, ambipolar diffusion,
ionization, and chemical reactions. Moreover, numerical modeling is necessary for self-consistent computations of the characteristics of clouds based
on observational data \cite{Pavlyuchenkov:2004}.

One promising direction is modeling the dust and molecular-line spectra and images of starless and protostellar cores based on modern dynamical models
(see, for example, \cite{Zhou1992, Rawlings1992}). Since the line parameters
(position, width, asymmetry) are sensitive to the velocity distribution inside the core, analysis of these lines can be used to determine the dynamical status of the core. At the same time, the molecular line emission depends strongly on many other parameters: the density distribution, temperature gradients, and concentrations of compound molecules \cite{Pavlyuchenkov2008}. This imposes additional requirements on the construction of dynamical models. A dynamical model for a starless or protostellar core should not only describe the evolution of the density and velocity distributions, but also adequately predict its thermal and chemical structure. So far, as a rule, modeling of the radiation spectra of protostellar clouds has been carried out using one-dimensional approximations. For instance, the evolution of the line profiles emitted by a protostellar cloud based on a one-dimensional dynamical model was studied in \cite{Keto2005}. Theoretical molecular-line spectra for a chemical-dynamical model of the pre-stellar core L1544 are presented and compared with observations in \cite{Pavlyuchenkov2003}. 

Our aim here is to construct a physical model for starless cores and use it to model their spectra. Our approach includes a description of the main physical
processes determining the dynamics of the core and the formation of the line profiles. Our dynamical model is two-dimensional and axially symmetrical, and takes into account the influence of rotation and magnetic fields, ohmic and ambipolar diffusion, ionization, heating and cooling processes, and radiative transfer. A non-LTE method is used to model the line radiation, which is able to describe both optically thin and optically thick line radiation. We emphasize that we rely on a complex approach that unifies a self-consistent dynamical model and a model for the radiative transfer in  molecular lines. Such an approach has been used only in a one-dimensional approximation \cite{Zhou1992, Rawlings1992, Keto2005}, which does  not take into account important physical processes associated with rotation and magnetic fields.

The article is organized as follows. Section 2 describes the main equations for the dynamical model and the numerical method used to solve them. Section 3 discusses the model for the radiative transfer in molecular lines that is used. Section 4 presents the results of numerical simulations of the structures of equilibrium and collapsing protostellar clouds. The Conclusion briefly discusses the main results of our study.

\section{Dynamical model}

\subsection{Basic equations}

We use the following system of equations to describe the dynamical evolution of magnetic, rotating protostellar clouds:
\begin{equation}\label{eq201}
  \pdiff{\rho}{t} + \nabla \cdot \left( \rho \vec{v} \right) = 0,
\end{equation}
\begin{equation}\label{eq202}
  \pdiff{\vec{v}}{t} + \left (\vec{v} \cdot \nabla \right) \vec{v} =
  -\fr{1}{\rho} \nabla P -
  \fr{1}{4\pi\rho}
  \left( \vec{B} \times \left( \nabla \times \vec{B} \right)\right) -
  \nabla \Phi,
\end{equation}
\begin{equation}\label{eq203}
  \pdiff{\vec{B}}{t} =
  \nabla \times \left(
  \vec{v} \times \vec{B} +
  \vec{v}_{\text{D}} \times \vec{B} -
  \eta_{\text{OD}} \left( \nabla \times \vec{B} \right)
  \right), \,\,
  \nabla \cdot \vec{B} = 0,
\end{equation}
\begin{equation}\label{eq204}
  \rho \left(
  \pdiff{\varepsilon}{t} + \left( \vec{v} \cdot \nabla \right) \varepsilon
  \right) + P \nabla \cdot \vec{v} =
  \Gamma_{\text{g}} - \Lambda_{\text{g}} - \Lambda_{\text{gd}},
\end{equation}
\begin{equation}\label{eq205}
  \nabla^2 \Phi = 4 \pi G \rho.
\end{equation}
Here, $\rho$ is the density, $\vec{v}$ the velocity, $P$ the pressure, $\varepsilon$ the specific internal energy, $\vec{B}$ the magnetic field, and  $\Phi$ the gravitational potential. The induction equation (\ref{eq203}) includes ohmic and ambipolar diffusion of the magnetic field. The ohmic diffusion coefficient is $\eta_{\text{OD}} = \sfr{c^2}{(4\pi\sigma)}$, where $\sigma$ is the conductivity of the plasma. We have assumed standard magnetic ambipolar diffusion (\cite{Spitzer1981}, Section 13.3), when the charged components of the plasma all move relative to the neutral components under the action of electromagnetic forces with essentially the same speed,
\begin{equation}\label{eq206}
  \vec{v}_{\text{D}} =
  \fr{\left(\nabla \times \vec{B}\right) \times \vec{B}}{4\pi R_{\text{pn}}},
\end{equation}
where $R_{\text{pn}}$ is the total coefficient of friction for the relative movement between the charged and neutral components. The energy equation (\ref{eq204}) includes heating and cooling processes: $\Gamma_{\text{g}}$ and $\Lambda_{\text{g}}$ are the gas heating and cooling rates per unit volume and $\Lambda_{\text{gd}}$ the rate at which energy is redistributed between the gas and dust via collisions per unit volume. The system (\ref{eq201}-\ref{eq205}) is closed with the equation of state of an ideal gas, $P = (\gamma - 1)\rho\varepsilon$, where $\gamma = 5/3$ is the
adiabatic index. This value for the adiabatic index is dictated by the fact that the rotational degrees of freedom of the H$_2$ molecules are frozen at the temperatures considered ($5~K \le T \le 100~K$) \cite{Landau1976}.

It is convenient to split the entire system (\ref{eq201}-\ref{eq205}) into simpler subsystems of equations according to the physical processes involved. The ideal magneto-gasdynamical equations can serve as the basic subsystem. Further, the Poisson equation for the gravitational potential, heating and cooling processes, radiative transfer, and ionization and diffusion of the
magnetic field can be grouped into subsystems. Each subsystem can be solved using the most appropriate numerical methods.

The dynamics of magnetic, rotating protostellar clouds can be studied in an axially symmetrical approximation if the initial magnetic field $\vec{B}$ is colinear with the angular velocity of the rotation $\vec{\Omega}$ and there are no azimuthal perturbations. In this case, it is convenient to use cylindrical coordinates $(r, \varphi, z)$. In the case of axial and equatorial symmetry, the numerical modeling can be carried out in the two-dimensional computational domain $(0 \le r,z \le R)$, where $R$ is the size of the computational domain. The radius of the cloud was specified to be less than $R$, with the boundary of the cloud moving together with the gas during the compression process. 

The conditions at the outer boundaries of the computational region $r = R$ and $z = R$ correspond to setting the derivatives of all computational quantities $f$ in the direction of the outward normal to the boundary $\vec{n}$ equal to zero: $\spdiff{f}{\vec{n}} = 0$. The condition of equatorial symmetry is imposed in the $z = 0$ plane, and the condition of axial symmetry at the $r = 0$ axis.

\subsection{Ionization balance equations} \label{ionsec}

To correctly describe the diffusion of the magnetic field in protostellar clouds, we must know the number densities of electrons $n_{\text{e}}$, ions $n_{\text{i}}$, and dust grains with charges of $-e$ and $+e$ ($n_{\text{g}_{-}}$ and $n_{\text{g}_{+}}$, in accordance with \cite{Nakano:2002}). We used an ionization model based on that of \cite{Tassis:2005}, which is relatively simple, but provides good agreement with the results of complex astrochemical computations such as those of
\cite{Shematovich:2003}. We included only one atomic ion, Mg$^+$, and one molecular ion, H$_3^+$, with number densities of $n_{\text{a}_{+}}$, $n_{\text{m}_{+}}$. The Mg$^+$ ion is formed in charge-transfer reactions
between atomic Mg and H$_3^+$, and also via photoionization by stellar radiation. The H$_3^+$ ion arises due to rapid reactions, beginning with the ionization of hydrogen and helium by cosmic rays, and the rate of production of H$_3^+$ is equal to the rate of ionization of hydrogen and helium, $\zeta = 3\cdot 10^{-17}~\text{s}^{-1}$ \cite{Webber:1998}. In the model used, the rates of ionization and recombination are sufficiently rapid that we can assume ionization equilibrium. Therefore, the equations for calculating $n_{\text{a}_{+}}$, $n_{\text{m}_{+}}$, $n_{\text{g}_{-}}$, and $n_{\text{g}_{+}}$ have the form
\begin{eqnarray}
  \left( 
  P_{\text{a}_0} + C_{\text{a}_0 \text{m}_+} n_{\text{m}_+} 
  \right) n_{\text{a}_0} = 
  \left( 
  C_{\text{a}_+ \text{e}} n_{\text{e}} + 
  C_{\text{a}_+ \text{g}_-} n_{\text{g}_-} + 
  C_{\text{a}_+ \text{g}_0} n_{\text{g}_0} 
  \right) n_{\text{a}_+} \;,
  \label{eq:ap}
  \\
  \zeta n_{\text{n}} = 
  \left( 
  C_{\text{dr}} n_{\text{e}} + 
  C_{\text{m}_+ \text{g}_-} n_{\text{g}_-} + 
  C_{\text{m}_+ \text{g}_0} n_{\text{g}_0} + 
  C_{\text{a}_0 \text{m}_+} n_{\text{a}_0} 
  \right) n_{\text{m}_+} \;,
  \label{eq:mp}
  \\
  C_{\text{eg}_0} n_{\text{e}} n_{\text{g}_0} = 
  \left(
  C_{\text{g}\pm} n_{\text{g}_+} + 
  \sum_\beta C_{\beta_+ \text{g}_-} n_{\beta_+}
  \right) n_{\text{g}_-} \;,
  \label{eq:gm}
  \\
  n_{\text{g}_0} \sum_\beta C_{\beta_+ \text{g}_0} n_{\beta_+} = 
  \left(
  C_{\text{g}\pm} n_{\text{g}_-} + 
  C_{\text{eg}_+} n_{\text{e}} 
  \right) n_{\text{g}_+} \;,
  \label{eq:gp}
\end{eqnarray}
where $P_{\text{a}_0}$ is the probability of ionization of an Mg atom by stellar radiation per second (\cite{Spitzer1981}, Section 5.2), $n_{\text{a}_0} = n_{\text{a}} - n_{\text{a}_+}$ the number density of neutral Mg atoms in the gas phase, $n_{\text{a}}$ the total number density of Mg in the gas phase, $n_{\text{n}}$ the number density of H, H$_2$ and He, and  $n_{\text{g}_0}$ the number density of neutral dust grains. The subscript $\beta$ denotes Mg and H$_3$. The notation for the reaction constants is indicated in the Table. The electron density is found from the condition of electrical neutrality. When calculating $P_{\text{a}_0}$, the photoionization cross section was set equal to the threshold value and taken out of the integral over frequency. The photorecombination coefficient $C_{\text{a}_+ \text{e}}$ was taken to be that for a hydrogen-like ion (\cite{Spitzer1981}, Section 5.1). We used the results of \cite{McCall:2004} to approximate the coefficient for dissociational recombination of H$_3^+$: $C_{\text{dr}} = 1.28 \cdot 10^{-6} T_{\text{g}}^{-1/2}~\text{cm}^3 \text{s}^{-1}$, where $T_{\text{g}}$ is the gas temperature.

\begin{table}[t]
\centering
\begin{tabular}{|l|l|c|} \hline
 Notation & Reaction & Reference \\ \hline
 $C_{\beta_0 \text{m}_+}$ & Charge transfer for H$_3^+$ and atoms $\beta$ & \cite{UMIST95} \\
 $C_{\beta_+ \text{e}}$ & Photorecombination of ions $\beta$ & \cite{Spitzer1981} \\
 $C_{\beta_+ \text{g}_-}$ & Recombination of ions $\beta$ on negative grains & \cite{Tassis:2005} \\
 $C_{\beta_+ \text{g}_0}$ & Recombination of ions $\beta$ on neutral grains & \cite{Tassis:2005} \\
 $C_{\text{dr}}$ & Dissociational recombination of H$_3^+$ & \cite{McCall:2004} \\
 $C_{\text{eg}_0}$ & Electron capture by neutral grains & \cite{Tassis:2005} \\
 $C_{\text{g}\pm}$ & Mutual neutralization of grains & \cite{Tassis:2005} \\
 $C_{\text{eg}_+}$ & Neutralization of positive grains & \cite{Tassis:2005} \\
 \hline
\end{tabular}
\caption{Notation and references for the ionization and recombination reaction constants}%
\label{t1}
\end{table}

In contrast to \cite{Tassis:2005} we included the freezing of Mg onto dust grains. We assumed that, after the recombination of Mg$^+$ on a dust grain, the kinetic energy of the formed Mg atom exceeds the adsorption energy, so that only collisions with neutral atoms and dust grains were taken into account when calculating the Mg adsorption rate \cite{Flower:2005}. At the gas and dust temperatures characteristic of protostellar clouds, the Mg desorption rate is much less than the Mg adsorption rate, so that the Mg abundance in the gas phase, $X_{\text{a}} = n_{\text{a}} / n_{\text{n}}$, is appreciably non-equilibrium. Neglecting the convective derivatives in the continuity equations for the neutrals, Mg$^+$, and Mg, and also equating the divergences of their velocities, we obtain an approximate equation that will be obeyed by $X_{\text{a}}$ at a fixed point:
\begin{equation}\label{eq:adsorb}
  \pdiff{}{t} \ln X_{\text{a}} =
  \left(\frac{n_{\text{a}_+}}{n_{\text{a}}} - 1\right) 
  P_{\text{s}} v_{\text{a}} s_{\text{g}} n_{\text{g}} \,
\end{equation}
where $P_{\text{s}} = 0.3$ is the probability of adsorption of an atom during a collision with a dust grain (see, for example, \cite{Shematovich:2003}),
$v_{\text{a}}$ the atom's mean thermal velocity, $s_{\text{g}}$ and $n_{\text{g}}$ the mean geometrical cross section and number density of the dust grains. We used a high initial Mg abundance in the gas phase: $X_{\text{a}}(0) = 5\cdot 10^{-7}$.

It is believed that the effective radii of dust grains have a power-law distribution with index $-3.5$ \cite{Spitzer1981}, with minimum and maximum radii of 10~nm and 1~$\mu$m. The ratio of the total concentrations of charged
and neutral dust grains to $n_{\text{n}}$ is constant and determined
from the mass fraction of dust, 1\%.

\subsection{Description of the numerical method}

\subsubsection{Magneto-gas dynamics}

The hyperbolic subsystem of equations separated out from the full system  (\ref{eq201}-\ref{eq205}) can be written in conservative form in cylindrical coordinates as follows:
\begin{equation}\label{eq207}
  \pdiff{\Vec{U}}{t} +
  \fr{1}{r}\pdiff{}{r}\left(r\Vec{F}\right) +
  \pdiff{\Vec{G}}{z} =
  \Vec{S},
\end{equation}
where the vectors of conservative variables $\Vec{U}$, fluxes $\Vec{F}$ and $\Vec{G}$, and sources $\Vec{S}$ are determined by the expressions
\begin{equation}\label{eq208}
  \Vec{U} =
  \left(
  \begin{array}{c}
   \rho \\
   \rho v_r \\
   \rho v_z \\
   \rho v_{\varphi} \\
   B_r \\
   B_z \\
   B_{\varphi} \\
   e
  \end{array}
  \right), \,\,
  \Vec{F} =
  \left(
  \begin{array}{c}
   \rho v_r \\
   \rho v_r^2 + P_{*} - \sfr{B_r^2}{4\pi} \\
   \rho v_r v_z - \sfr{B_r B_z}{4\pi} \\
   \rho v_r v_{\varphi} - \sfr{B_r B_{\varphi}}{4\pi} \\
   0 \\
   v_r B_z - v_z B_r \\
   v_r B_{\varphi} - v_{\varphi} B_r \\
   v_r (e + P_{*}) - \sfr{B_r (\vec{v} \cdot \vec{B})}{4\pi}
  \end{array}
  \right),
\end{equation}
\begin{equation}\label{eq209}
  \Vec{G} =
  \left(
  \begin{array}{c}
   \rho v_z \\
   \rho v_z v_r - \sfr{B_z B_r}{4\pi} \\
   \rho v_z^2 + P_{*} - \sfr{B_z^2}{4\pi} \\
   \rho v_z v_{\varphi} - \sfr{B_z B_{\varphi}}{4\pi} \\
   v_z B_r - v_r B_z \\
   0 \\
   v_z B_{\varphi} - v_{\varphi} B_z \\
   v_z (e + P_{*}) - \sfr{B_z (\vec{v} \cdot \vec{B})}{4\pi}
  \end{array}
  \right), \,\,
  \Vec{S} =
  \left(
  \begin{array}{c}
   0 \\
   \left(\rho v_{\varphi}^2 + P_{*} - \sfr{B_{\varphi}^2}{4\pi}\right) / r -
   \rho \spdiff{\Phi}{r} \\
   - \rho \spdiff{\Phi}{z} \\
   \left( -\rho v_r v_{\varphi} + \sfr{B_r B_{\varphi}}{4\pi} \right)/r \\
   0 \\
   \left(v_z B_r - v_r B_z\right)/r \\
   0 \\
   -\rho (\vec{v} \cdot \nabla)\Phi
  \end{array}
  \right),
\end{equation}
where $P_{*} = P + \sfr{\vec{B}^2}{8\pi}$ is the total pressure and, $e = \rho
\left(\varepsilon + \sfr{\vec{v}^2}{2}\right) + \sfr{\vec{B}^2}{8\pi}$ the total energy density.

During the compression, there is an appreciable concentration of matter in the central part of the cloud, leading to strong radial gradients in the density,
velocity, and other quantities. Therefore, to enhance the accuracy of the computations, we used an adaptive moving mesh in our numerical code, which became denser toward the center with time \cite{Dudorov2003b}. The system (\ref{eq207}-\ref{eq209}), was approximated in a moving curvilinear coordinate system that was related to the initial coordinate system by the transformations $t = \tau$, $r = r(\tau, \xi)$, $z = z(\tau, \zeta)$. The initial equations (\ref{eq207}-\ref{eq209}) in the new variables $\tau$, $\xi$, and $\zeta$ have the form
\begin{equation}\label{eq210}
  \pdiff{}{\tau}\left(r Q_r Q_z \Vec{U}\right) +
  \pdiff{}{\xi}\left(r Q_z \tilde{\Vec{F}}\right) +
  \pdiff{}{\zeta}\left(r Q_r \tilde{\Vec{G}}\right) =
  r Q_r Q_z \Vec{S},
\end{equation}
where $\tilde{\Vec{F}} = \Vec{F} - \omega_r \Vec{U}$, $\tilde{\Vec{G}} = \Vec{G} - \omega_z \Vec{U}$, $\omega_{r,z}$ is the velocity of the new coordinate system relative to the initial system, and $Q_{r,z}$ are metric coefficients. A more detailed description of the technique used to construct
the adaptive mesh is presented in the Appendix. In the computational results presented below, the size of a central cell was approximately 0.1~AU, while the
radius of the cloud itself was 0.2 pc.

We introduced in the moving curvilinear coordinate system $\xi$, $\zeta$ in the computational domain a uniform difference mesh whose structure was determined by the distribution of the mesh coordinates $\xi_i = i \Delta \xi$, $\zeta_j = j \Delta \zeta$, where $\Delta\xi = R/N_r$, $\Delta\zeta = R/N_z$, and the subscripts $i = 0, 1, \ldots, N_r$ and $j = 0, 1, \ldots, N_z$. The
computational mesh quantities $\Vec{U}^n_{c}$ at time $\tau^n$ correspond
to the cells $c = (i + 1/2, j + 1/2)$, which are numbered with half-integer indexes. The values of the mesh quantities
\begin{equation}\label{eq211}
  \Vec{U}_c^n =
  \fr{1}{(r Q_{r} Q_{z})_c^n \Delta\xi \Delta\zeta}
  \int\limits_{\xi_1}^{\xi_2} d\xi
  \int\limits_{\zeta_1}^{\zeta_2} d\zeta
  (r Q_{r} Q_{z} \Vec{U})^n,
\end{equation}
where $r_c = \sfr{(r_1 + r_2)}{2}$, at time $\tau^{n+1} = \tau^n +
\Delta \tau$ were computed using the difference scheme
\begin{equation}\label{eq212}
  \fr{
  \left(r Q_r Q_z \Vec{U}\right)_c^{n+1} -
  \left(r Q_r Q_z \Vec{U}\right)_c^n
  }{\Delta\tau} +
  \fr{\Delta_{\xi}\left(r Q_z \tilde{\Vec{F}}\right)}{\Delta\xi} +
  \fr{\Delta_{\zeta}\left(r Q_r \tilde{\Vec{G}}\right)}{\Delta\zeta} =
  \left(r Q_r Q_z \Vec{S}\right)_c^n.
\end{equation}
The components of the source vector $\Vec{S}_c^n$ were determined using  expressions (\ref{eq209}), which were computed at the center of cell $c$ at time $\tau^n$. The operators $\Delta_{\xi}$ and $\Delta_{\zeta}$ determine the difference between quantities at the cell boundaries in the corresponding directions.

We computed the numerical vector fluxes $\Vec{F}$ and $\Vec{G}$ through the cell boundaries by splitting them along spatial directions. The numerical fluxes in each spatial direction were computed based on the solution of the correspoding one-dimensional Riemann problem for the decay of an arbitrary discontinuity. The split one-dimensional Riemann problem can be written in the form (we consider only the $\xi$ direction as an example):
\begin{equation}\label{eq213}
  \pdiff{\Vec{U}}{\tau} + \pdiff{\tilde{\Vec{F}}}{\xi} = 0, \ \
  \Vec{U}(\xi, 0) =
  \left \{
  \begin{array}{ll}
   \Vec{U}_L, & \xi < 0, \\
   \Vec{U}_R, & \xi > 0.
  \end{array}
  \right.
\end{equation}
Let us consider the solution of this problem in the approximation when all propagating waves that arise after the decay of the initial arbitrary discontinuity represent strong discontinuities. The corresponding Hugoniot conditions should be satisfied at each such discontinuity:
\begin{equation}\label{eq214}
  \tilde{D}_{\alpha}[\Vec{U}]_{\alpha} =
  [\Vec{F} - \omega_{\alpha} \Vec{U}]_{\alpha},
\end{equation}
where $\alpha$ denotes the index of a discontinuity with velocity  $\tilde{D}_{\alpha}$ and the square brackets denote the difference
between the right-hand and left-hand quantities for the given discontinuity. Note that, in this approximation for the solution of the Riemann problem, some
propagating waves can represent non-evolving rarefaction shocks.

The numerical fluxes $\tilde{\Vec{F}}$ through the cell boundaries in the scheme (\ref{eq212}) are determined as follows:
\begin{equation}\label{eq215}
  \tilde{\Vec{F}}_{i} =
  \fr{\Vec{F}_{L} + \Vec{F}_{R}}{2} -
  \fr{1}{2}
  \sum\limits_{\alpha}|\tilde{D}_{\alpha}|[\Vec{U}]_{\alpha} +
  \delta\tilde{\Vec{F}}_{i},
\end{equation}
where
\begin{equation}\label{eq216}
  \delta \tilde{\Vec{F}}_{i} =
  \fr{1}{2}
  \sum\limits_{\alpha+}
  {\rm limiter}_{\alpha}
  \left(
   \Delta\tilde{\Vec{F}}^{\alpha+}_{i - 1},
   \Delta\tilde{\Vec{F}}^{\alpha+}_{i}
  \right) -
  \fr{1}{2}
  \sum\limits_{\alpha-}
  {\rm limiter}_{\alpha}
  \left(
   \Delta\tilde{\Vec{F}}^{\alpha-}_{i},
   \Delta\tilde{\Vec{F}}^{\alpha-}_{i + 1}
  \right),
\end{equation}
\begin{equation}\label{eq217}
  \Delta\tilde{\Vec{F}}^{\alpha\pm}_{i} =
  \fr{\tilde{D}_{\alpha} \pm |\tilde{D}_{\alpha}|}{2}
  [\Vec{U}]_{\alpha}.
\end{equation}
The sum in the first term in (\ref{eq216}) is carried out over all positive velocities $\tilde{D}_{\alpha}$, and the sum in the second term over all negative velocities. The function ${\rm limiter}(x,y)$ restricts the values of the anti-diffusion corrections $\delta\tilde{\Vec{F}}$ to ensure that the scheme is monotonic. We used the following non-linear limiting function for all $\alpha$ \cite{Chakravarthy1985, Dudorov1999a}:
\begin{equation}\label{eq218}
  {\rm limiter}(x, y) =
  \fr{1 + p}{2} {\rm minmod}(y, q x) +
  \fr{1 - p}{2} {\rm minmod}(q y, x),
\end{equation}
where
\begin{equation}\label{eq219}
  {\rm minmod}(x, y) =
  \fr{1}{2}
  \left[{\rm sign}(x) + {\rm sign}(y)\right]
  \min(|x|, |y|).
\end{equation}
When $p = 1/3$ in regions where the solution is smooth, this scheme has third-order accuracy in the spatial variable and first-order accuracy in time. The value of $q$ was equal to 4 in all our computations \cite{Dudorov1999a}.

In our numerical computations, we used a difference scheme based on solving the Riemann problem in the HLLD approximation \cite{Miyoshi2006}, which assumes
that five strong discontinuities with four intermediate states are formed during the decay of the initial discontinuity: two MHD shocks, two Alfven
(rotational) discontinuities, and an MHD contact (tangential) discontinuity. A generalized Lagrange multiplier (GLM) method was used to clean the divergence of the magnetic field \cite{Dedner2002}. The described difference scheme for approximating the equations in the hyperbolic system (\ref{eq201}-\ref{eq205}) is a total variation diminishing (TVD) scheme with third-order accuracy in the spatial variable in smooth regions of the solution and first-order accuracy in time. With the corresponding boundary conditions, the difference scheme approximating the ideal magneto-gasdynamical equations exactly conserves the total mass and energy. The stability of the scheme is provided by limiting the time step $\Delta \tau$ (the Courant-Friedrichs-Levy condition) \cite{Dudorov1999a}. The Poisson equation for the gravitational potential was solved using ADI method \cite{Dudorov1999b}.

To verify the computational properties and accuracy of the hyperbolic part of the numerical code, we carried out a series of test computations for
problems with known exact or approximate analytical solutions (for details, see \cite{Dudorov1999a, Dudorov1999b}). Moreover, we checked the correctness of the operation of the code using self-similar solutions describing the inhomogeneous, free, and isothermal collapse of protostellar clouds \cite{Dudorov2008}. The second self-similar solution corresponds to the critical case of the propagation of a rarefaction shock near the time when it is focused in the central part of the collapsing protostellar cloud. These computational results demonstrated the good computational properties of the code and its suitability for a wide range of astrophysical problems.

\subsubsection{Thermal Structure}

We separated out a subsystem of equations to describe the thermal structure of a protostellar cloud:
\begin{eqnarray}
  \rho\dfrac{\partial \varepsilon}{\partial t} =
  \Gamma_{\text{g}} - \Lambda_{\text{g}} -
  \Lambda_{\text{gd}}, \label{therm01} \\
  \Gamma_{\text{d}} - \Lambda_{\text{d}} +
  \Lambda_{\text{gd}} = 0, \label{therm02}
\end{eqnarray}
where $\varepsilon$ is the specific internal energy of the gas, $\Gamma_{\text{g}}$ and $\Gamma_{\text{d}}$ the gas and dust heating rates per unit volume, and $\Lambda_{\text{g}}$ and $\Lambda_{\text{d}}$ the gas and dust cooling rates per unit volume.  Equations \eqref{therm01} and \eqref{therm02} are interrelated by the rate of the redistribution of energy between the gas and dust due to collisions between gas molecules and dust grains, $\Lambda_{\text{gd}}$. Thus, the temperature distributions for the dust and gas must be found jointly with the solution of this system. The estimates of \cite{Leger1985} show that, given the relatively low heat capacity of the dust, the time required to establish radiative equilibrium for the dust component is much shorter than the characteristic dynamical time-scale for a wide range of densities and temperatures in the models considered. Therefore, we neglected the time derivative of the thermal energy of the dust in~\eqref{therm02}; i.e., we assumed that the dust was in radiative equilibrium at each time. Further, we describe the heating and cooling processes that must be included when modeling the cores of molecular clouds.

One of the main mechanisms for heating the gas, especially in the central, dense regions of the cloud, is heating by cosmic rays. We used the following formula from~\cite{Falgarone:1985}:
\begin{equation}
  \Gamma_{\text{cr}} = 10^{-27}\, n(\text{H}_2),
\end{equation}
where $n(\text{H}_2)$ is the number density of molecular hydrogen. However, the intensity of the cosmic-ray flux is rather uncertain, and can vary appreciably for different clouds (see, for example, \cite{Lintott2006}). An important gas heating mechanism in the envelope is the photoelectric effect~\cite{Bakes:1994, Weingartner:2001}, when high-energy interstellar UV photons eject electrons from the surface of dust grains. These electrons have  comparatively high energies, and so heat the gas. To calculate the heating function for the photoelectric effect, we used formula (42) from~\cite{Bakes:1994}:
\begin{equation}
  \Gamma_{\text{pe}} = 10^{-24}\,
  \epsilon_{\text{pe}}\, G_0\,  n(\text{H}),
\end{equation}
where $\epsilon_{\text{pe}}$ is the efficiency of photoelectric heating, $n(\text{H})$ the density of hydrogen (atomic and molecular), and $G_0$ the number of high-energy photons (6~eV $<$ $h\nu$ $<$ 13.6~eV) at a given point normalized to the mean number of high-energy photons in the interstellar radiation field~(see~\cite{Weingartner:2001}, Section~4.1).

We calculated $\epsilon_{\text{pe}}$ using formula (43) from~\cite{Bakes:1994}, taking the number density of free electrons in this formula from the ionization model described in Section~\ref{ionsec}. In our computations, $\epsilon_{\text{pe}}$ varied from 0.05 at the center of the cloud to 0.01 in the cloud envelope.

The gas in the cloud is primarily cooled by molecular-line radiation and collisions between the molecular gas and dust grains, which, as a rule, have lower temperatures. We used the following approximate formula from~\cite{Goldsmith:2001} to compute the molecular-line cooling function:
\begin{equation} \label{mol}
  \Lambda_{\text{ml}} = \alpha \left(\dfrac{T_\text{g}}{10\,K}\right)^\beta,
\end{equation}
where $T_{\text{g}}$ is the gas temperature and the coefficients $\alpha$ and $\beta$ depend on the density (see~\cite{Goldsmith:2001}, Table~2). This formula generalizes results of modeling the radiative transfer in a molecular cloud and of computing the cooling functions for cooling in lines of CO, $^{13}$CO, C, C$^{+}$ and other compound molecules. The rate of energy exchange between the gas and dust was calculated using formula (15) from~\cite{Goldsmith:2001}:
\begin{equation} \label{gd}
  \Lambda_{\text{gd}} = 6.32\cdot 10^{-34}\,
  n^2(\text{H}_2)\,T_{\text{g}}^{1/2}\,
  (T_{\text{g}} - T_{\text{d}}),
\end{equation}
where $T_{\text{d}}$ is the dust temperature. Depending on the ratio of $T_{\text{d}}$ and $T_{\text{g}}$, the energy exchange between the gas and dust can lead to either cooling and heating of the gas.

The main heating mechanism for the dust is the absorption of radiation:
\begin{equation} \label{gamma_d}
  \Gamma_{\text{d}} = 4\pi \int\limits_{0}^{\infty}
  \rho_{\text{d}}\, \kappa_{\nu} J_{\nu}\, d\nu,
\end{equation}
where $\rho_{\text{d}}$ is the mass of dust per unit volume, $\kappa_{\nu}$ the absorption coefficient per unit mass of dust, and 
\begin{equation} \label{Jnu}
  J_{\nu} = \dfrac{1}{4\pi} \int\limits_{4\pi} I_{\nu} d\Omega
\end{equation}
the angle-averaged spectral intensity of the radiation. The dust is cooled primarily by its own blackbody (thermal) radiation:
\begin{equation}  \label{lambda_d}
  \Lambda_{\text{d}} = 4\pi \int\limits_{0}^{\infty}
  \rho_{\text{d}}\, \kappa_{\nu} B_{\nu}(T_{\text{d}})\, d\nu,
\end{equation}
where $B_{\nu}(T_{\text{d}})$ is the Planck function for the dust temperature $T_{\text{d}}$. Computation of the mean intensity~\eqref{Jnu} requires knowledge of the spectral intensity, which can be obtained from the solution of the radiative transfer equation:
\begin{equation} \label{radtran}
  \dfrac{dI_{\nu}}{ds} = -\rho_{\text{d}} \kappa_{\nu} I_{\nu} +
  \rho_{\text{d}} \kappa_{\nu} B_{\nu}(T_{\text{d}}),
\end{equation}
where the first term on the right-hand side describes absorption of the radiation and the second term describes the blackbody radiation of the dust along a specified direction. Thus, determining the temperature of the dust requires a consideration of the system \eqref{therm02}, \eqref{gamma_d}-\eqref{radtran} describing the condition for radiative equilibrium with an internal source, $\Lambda_{\text{gd}}$. We now describe the method used to solve this system for the conditions characteristic of starless cores.

\begin{figure}[t]
\centering
\includegraphics[width=0.4\textwidth]{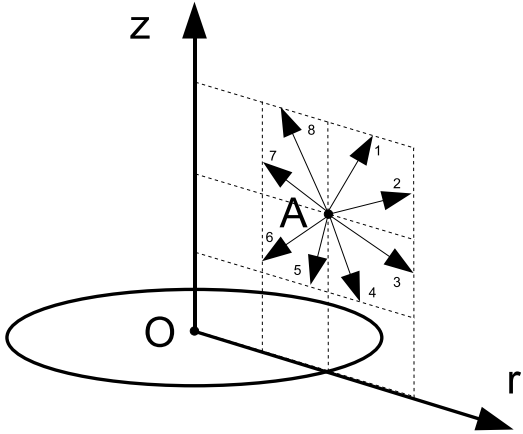}
\caption{Scheme for integrating the radiative-transfer equation. The mean intensity is calculated using the short-characteristic method, where the rays along which the integration is carried out lie in the meridional plane. The integration in each direction is carried out to cell's boundary, where the intensity is found via interpolation.}
\label{fig01}
\end{figure}

First and foremost, we assume that the cloud is heated primarily by external (interstellar) radiation, and that the cloud itself is optically thin to its own (thermal) radiation; i.e., we neglect the second term in~\eqref{radtran}:
\begin{equation} \label{reduce}
  \dfrac{dI_{\nu}}{ds} = -\rho_{\text{d}} \kappa_{\nu} I_{\nu}.
\end{equation}
This equation describes the weakening of the interstellar radiation due to its propagation through the cloud. This approximation is well satisfied for molecular cloud cores, and substantially simplifies the solution procedure. Further, we approximate the integral~\eqref{Jnu} in the form
\begin{equation} \label{Jsum}
  J_{\nu} = \dfrac{1}{N} \sum\limits_{i = 1}^{N} I_{i}(\nu),
\end{equation}
i.e., we consider a discrete set of $N$ directions for which the intensities $I_{i}(\nu)$ are determined. An important aspect of the organization of the computations is the choice of these directions. Since we are considering an axially symmetrical problem, we chose directions lying in the meridional plane (Fig.~\ref{fig01}).

Of course, this choice introduces some small error, however it appreciably simplifies the solution procedure, since we need not consider curvilinear trajectories during the integration. We used the short characteristic method to solve the transfer equation for $I_{i}(\nu)$~\cite{Kunasz:1988}. In this method, the integration~\eqref{reduce} in each cell is carried out from its center to its boundary, with the intensity at the cell boundary found via interpolation. To model the radiative transfer, we used the frequency dependence of the opacity $\kappa_{\nu}$ for compact, silicate dust grains with ice mantles from~\cite{Ossenkopf:1994}.

We tested the described method for modeling the thermal structure of the cloud using a series of model problems. In particular, we compared our computational results with the solution obtained by the TRANSPHERE-1D\footnote{http://www.mpia.de/homes/dullemon/radtrans} numerical code developed by C. Dullemond to model radiative transfer in dust envelopes taking into account the thermal radiation of the dust. The maximum difference in the dust temperatures in the obtained solutions was less than 0.2~K for all conditions considered, justifying our use of the approximation~\eqref{reduce}.

\begin{figure}[t]
\centering
\includegraphics[width=0.45\textwidth]{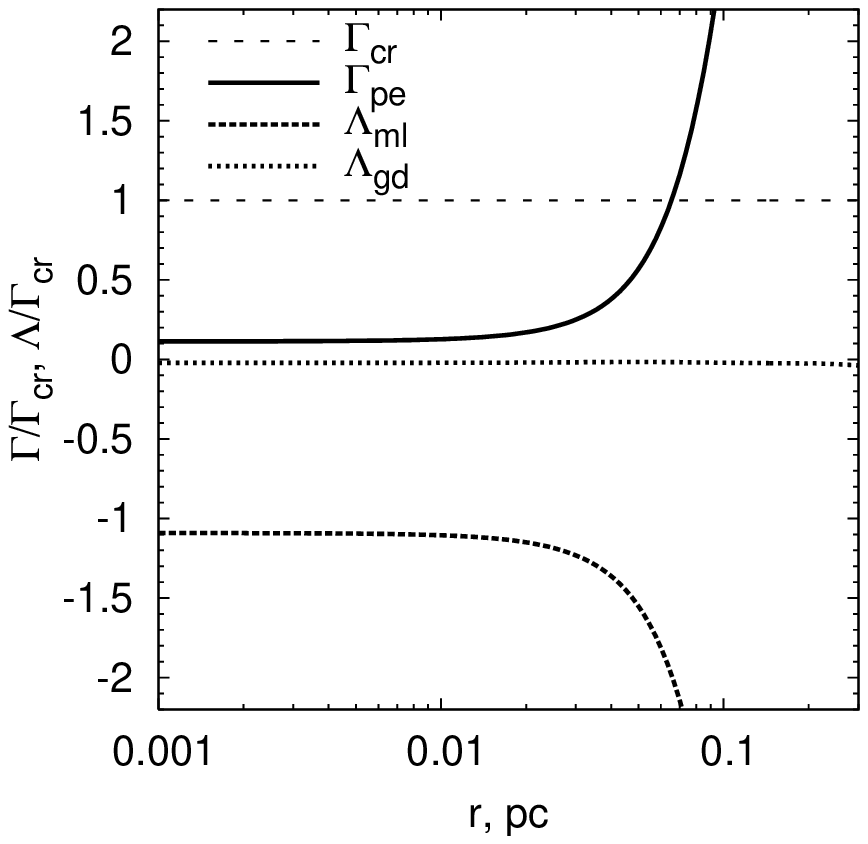}
\includegraphics[width=0.45\textwidth]{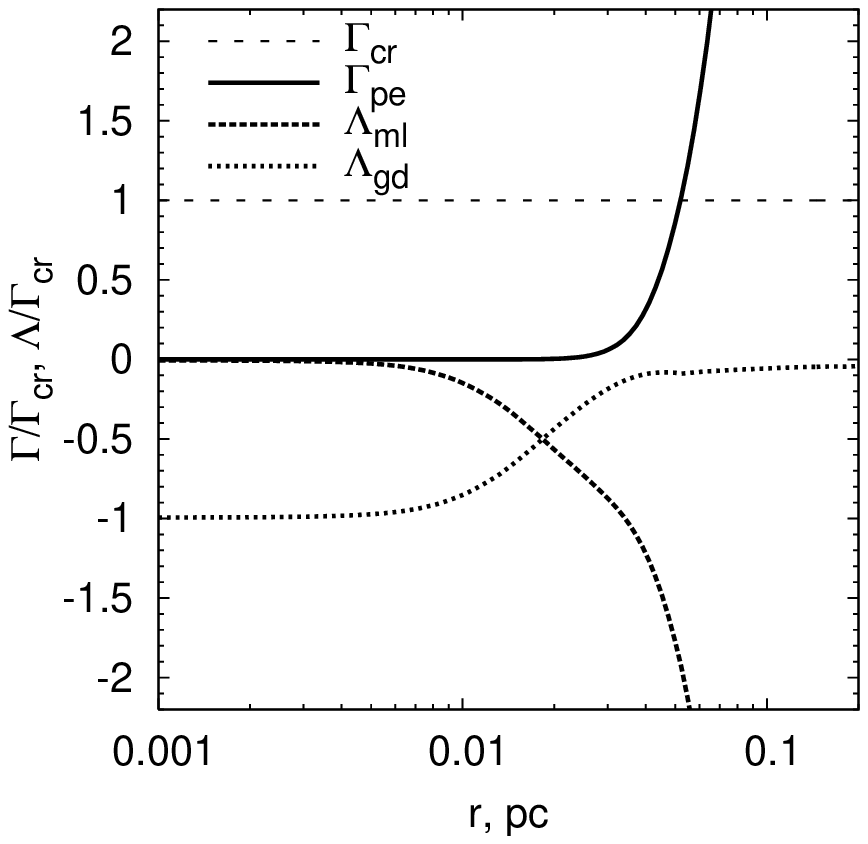}
\caption{The ratio of the heating/cooling rates of the gas to the heating rate by cosmic rays for equilibrium models of protostellar clouds with central densities $n_{\text{c}} = 10^4$~cm$^{-3}$ (left) and $n_{\text{c}} = 10^6$~cm$^{-3}$ (right)}
\label{fig02}
\end{figure}

To demonstrate the relative contributions of various processes to the thermal balance of the cloud, we present the ratio of heating and cooling rates of the gas to the heating function by cosmic rays for the equilibrium cloud model described in detail in Section~\ref{EquSect}. Figure~\ref{fig02} shows our results for models with central densities $n_{\text{c}} = 10^4$~cm$^{-3}$ and $n_{\text{c}} = 10^6$~cm$^{-3}$.

In the model with $n_{\text{c}} = 10^4$~cm$^{-3}$, the heating by cosmic rays and the photoelectric effect are comparable in magnitude in the central regions of the cloud, while photoelectric heating dominates in the cloud envelope. Cooling of the gas due to collisions with dust is appreciably lower than cooling via molecular-line radiation in the entire volume of the cloud. The dominant heating mechanisms in the cloud center and envelope in the model with $n_{\text{c}} = 10^6$~cm$^{-3}$ are clearly distinguished: cosmic rays and the photoelectric effect, respectively. The cooling mechanisms are also clearly distinguished: collisions with dust in the cloud center and molecular line radiation in the envelope.

\subsubsection{Diffusion of the magnetic field.}
\label{madsec}

To describe the diffusion of the magnetic fields in collapsing protostellar clouds, we can neglect the friction between the charged components of the plasma \cite{Nakano:2002}. We therefore assume that a component of type $\alpha$, consisting of particles with mass $m_\alpha$, charge $\pm e$, and number density $n_\alpha$ has conductivity $\sigma_\alpha = e^2 n_\alpha^2/R_{\alpha\text{n}}$, where $R_{\alpha\text{n}} =
\mu_{\alpha\text{n}} n_\alpha n_{\text{n}} \langle sv
\rangle_{\alpha\text{n}}$~is the coefficient of friction with neutral particles, $\mu_{\alpha\text{n}}$~the reduced mass, and $\langle sv
\rangle_{\alpha\text{n}}$~the deceleration coefficient. Dividing the charged particles into four types --- electrons, ions, and positive and negative dust
particles --- we can write the total conductivity of the plasma $\sigma =
\sigma_{\text{e}} + \sigma_{\text{i}} + \sigma_{\text{g}_{+}} +
\sigma_{\text{g}_{-}}$, and the total coefficient of friction between the charged and neutral particles $R_{\text{pn}} = R_{\text{en}} +
R_{\text{in}} + R_{\text{g}_{+}\text{n}} + R_{\text{g}_{-}\text{n}}$. For the ions, $\langle sv \rangle_{\text{in}} = 2\cdot 10^{-9}~\text{cm}^3\cdot\text{s}^{-1}$ (\cite{Spitzer1981}, Section 2.2). When computing the deceleration coefficient for the electrons, $\langle sv \rangle_{\text{en}}$ we used the approximate tabulated data of (\cite{PhysVal:1991}, p. 393). We assume for the charged dust grains
$\langle sv \rangle_{\text{g}_{+}\text{n}} = \langle sv \rangle_{\text{g}_{-}\text{n}} = s_{\text{g}} v_{\text{n}}$ ~\cite{Tassis:2005}, where $v_{\text{n}}$~is the mean thermal velocity of the neutral components.

Using (\ref{eq206}) the diffusion part of the induction equation (\ref{eq203}) can be rewritten:
\begin{equation}\label{eqMD1}
  \pdiff{\vec{B}}{t} =
  \eta \nabla^2\vec{B} +
  \nabla \times \left(u_{\text{AD}}\vec{B}\right) -
  \nabla\eta \times (\nabla \times \vec{B}),
\end{equation}
where $\eta = \eta_{\text{OD}} + \sfr{\vec{B}^2}{(4\pi R_{\text{pn}})}$ is the total diffusion coefficient and $u_{\text{AD}} = \sfr{(\vec{B} \cdot (\nabla
\times \vec{B}))}{(4\pi R_{\text{pn}})}$ the scalar ambipolar diffusion velocity. Since the diffusion coefficient $\eta$ and the scalar ambipolar diffusion velocity $u_{\text{AD}}$ depend on the magnetic field, the use of explicit schemes to solve (\ref{eqMD1}) leads to strong restrictions on the time step. Therefore, to approximate (\ref{eqMD1}) in the interval $\tau^{n} \le \tau \le \tau^{n+1}$ in the two-dimensional numerical code, we used an absolutely stable, fully implicit scheme, which was realized via the following
iterative process:
\begin{equation}\label{eqMD2}
  \fr{\vec{B}^{(p+1)} - \vec{B}^{(0)}}{\Delta\tau} =
  \eta^{(p)} \nabla^2\vec{B}^{(p+1)} +
  \nabla \times \left(u_{\text{AD}}^{(p)}\vec{B}^{(p)}\right) -
  \nabla\eta^{(p)} \times (\nabla \times \vec{B}^{(p)}),
\end{equation}
where $p$ is the iteration number. The values obtained via the solution of the hyperbolic subsystem of equations are used as the initial conditions for
$\vec{B}^{(0)}$. A splitting according to spatial directions (a locally one-dimensional scheme) is carried out in (\ref{eqMD2}), and all differential operators are approximated with finite differences with second order accuracy. The resulting system of linear algebraic equations was solved with Thomas algorithm along each spatial direction.

The ionization balance equations (\ref{eq:ap}-\ref{eq:gp}) were solved numerically using the Newton method. The resulting system of algebraic equations was solved using the method of Gauss.

\section{Model for radiative transfer in molecular lines}

As a rule, the formation of line radiation in molecular cloud cores occurs under non-LTE conditions (see, for example, \cite{Pavlyuchenkov2008}). Therefore, modeling these lines requires the consideration of a system of balance equations for the level populations of the molecules, together with the radiative-transfer equation. We solved this problem using the approach
described in detail in \cite{Pavlyuchenkov:2004}. Here, we present only the main ideas of this method.

The modeling of line radiation includes two stages. First, the distribution of the level populations (excitation temperatures of the transitions) of the molecules considered consistent with the external radiation and thermal radiation of the cloud was found. Accelerated $\Lambda$ iterations were used for it, where the mean intensity in each cell of the computational domain was found using the long-characteristic method. The idea behind the accelerated $\Lambda$ iterations is to use the radiation field computed in the previous iteration step to find a new approximation for the excitation temperature, which, in turn, is used to compute the intensity in the following iteration.
The rays along which the integration of the transfer equation is carried out were chosen using the Monte-Carlo method. The integration of the radiative-transfer equation took into account the continuum radiation of the dust, with the dust temperature taken to be specified.

In the second stage of the modeling, the obtained excitation-temperature distribution for the transitions was used to compute the observed properties
of the modeled object, namely the line profiles, spectral maps, integrated intensity maps, mean velocity maps, and so forth. The orientation of the modeled cloud in space can be chosen, as well as the parameters of the radio-telescope beam. Our numerical code also enables a detailed analysis of
the formation of a studied line. The modeling results were used to calculate the contribution function which provides a relative input of all elements of the cloud to the resulting line profile \cite{Pavlyuchenkov2008}. Such analyses represent a powerful tool for interpreting the modeling results.

The radiative transfer in molecular lines was modeled using a dedicated spatial grid that differs from that used to compute the dynamical evolution of the cloud. This is due to the fact that the criteria for the radiative transfer grid differ from those required for the hydrodynamical computations. Therefore, to construct the spatial grid for the line modeling we interpolated values from the hydrodynamical grid. The molecular data, such as the energy levels, Einstein coefficients, and collisional-excitation coefficients, were taken from the Leiden LAMBDA database\footnote{http://www.strw.leidenuniv.nl/$\sim$moldata/}~\cite{Schoier:2005}.
The frequency dependence of the opacity used for the dust radiation was the same as that used when modeling the thermal structure of the cloud. The code was well tested, and successfully used in the past to model radiative transfer in molecular cores~\cite{Pavlyuchenkov:2006, Pavlyuchenkov2008} and protoplanetary disks~\cite{Pavlyuchenkov:2007, Semenov:2008}.

\section{Demonstrative computations}

\subsection{Equilibrium configurations of protostellar clouds}
\label{EquSect}

The equilibrium structure of a protostellar cloud can be described using the system of equations:
\begin{equation}\label{eq401}
  \nabla P = -\rho\nabla\Phi, \,\,
  \nabla^2\Phi = 4\pi G \rho, \,\,
  P = \sfr{\mathcal{R}\rho T_{\text{g}}}{\mu},
\end{equation}
\begin{equation}\label{eq403}
  \Gamma_{\text{g}} - \Lambda_{\text{g}} - \Lambda_{\text{gd}} = 0,\,\,
  \Gamma_{\text{d}} - \Lambda_{\text{d}} + \Lambda_{\text{gd}} = 0,
\end{equation}
where $\mathcal{R}$ is the universal gas constant and $\mu$ the mean molecular weight. These equations were solved numerically in a spherically symmetrical approximation, iterating in the temperature of the gas and dust until the required accuracy was attained. The equations of hydrostatics (\ref{eq401}) and thermal balance of the gas and dust (\ref{eq403}) were solved in each iteration. The thermal-balance equation was solved using the Newton method, jointly with the radiative-transfer equation (\ref{reduce}) and the ionization-balance equations (\ref{eq:ap}-\ref{eq:gp}). The densities at the cloud center and in the outer envelope were specified as boundary conditions.
The radius and mass of the cloud were determined from the solution of (\ref{eq401}, \ref{eq403}).

\begin{figure}[t]
\centering
\includegraphics[width=0.45\textwidth]{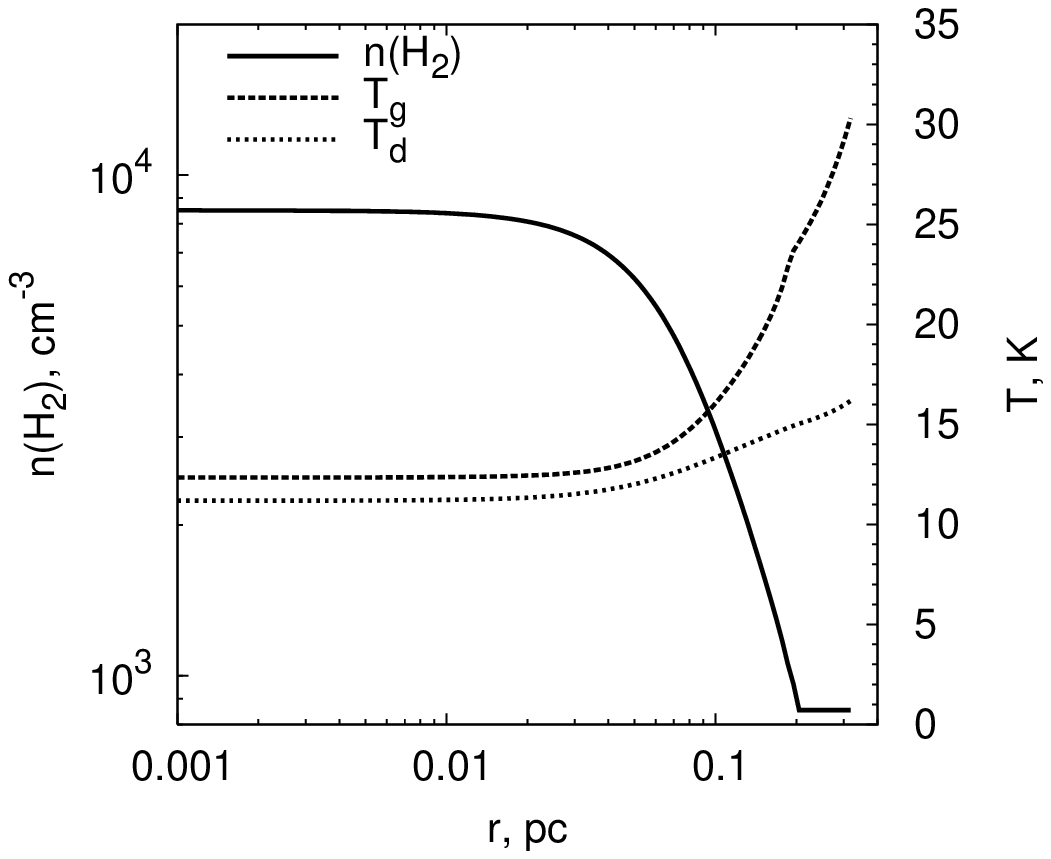}
\includegraphics[width=0.45\textwidth]{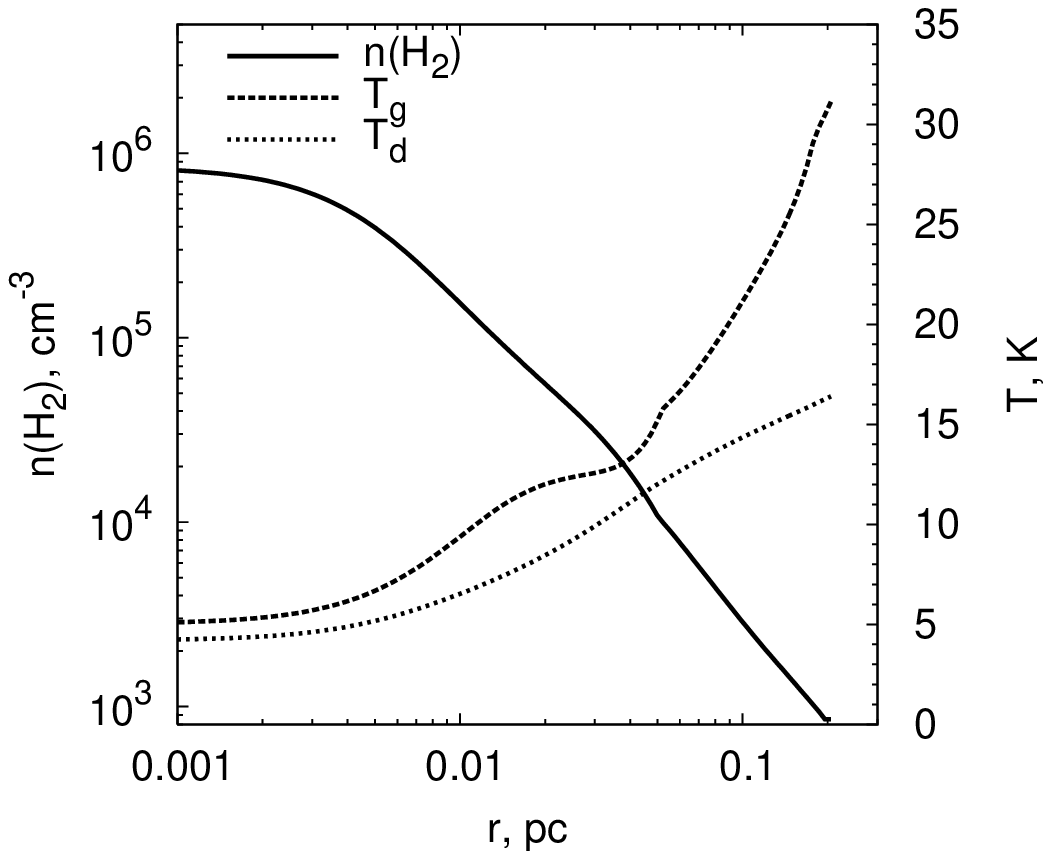}
\caption{Distributions of the number density of molecular hydrogen, the gas temperature, and the dust temperature for equilibrium cloud models with central densities $n_{\text{c}} = 10^4~\text{cm}^{-3}$ (left) and $n_{\text{c}} = 10^6~\text{cm}^{-3}$ (right).}
\label{fig03}
\end{figure}

Figure \ref{fig03} shows the distributions of the number density of molecular hydrogen $n(\text{H}_2)$, the dust temperature, and the gas temperature for equilibrium cloud models with central densities $n_{\text{c}} = 10^4$~cm$^{-3}$ (left) and $n_{\text{c}} = 10^6$~cm$^{-3}$ (right). The envelope density is $n_{\text{ext}} = 10^{3}$~cm$^{-3}$ in both models.

The radius and mass of the cloud core in the first model are $R_{\text{core}} = 0.2~\text{pc}$ and $M_{\text{core}} = 4.4~M_{\odot}$. The temperatures of the gas and dust at the cloud center are $T_{\text{g}} = 12.3~\text{K}$, $T_{\text{d}} = 11~\text{K}$, while the temperatures in the outer envelope are $T_{\text{g}} = 33.1~\text{K}$, $T_{\text{d}} = 16.9~\text{K}$. The difference in the gas and dust temperatures in the central parts of the cloud is due to the cooling of the gas in the cloud via molecular line radiation (left panel in Fig. \ref{fig02}). 

In the second model, $R_{\text{core}} = 0.2~\text{pc}$ and $M_{\text{core}} =
4.6~M_{\odot}$. The gas and dust temperatures at the cloud center are $T_{\text{g}} = 5~\text{K}$ and $T_{\text{d}} = 4.1~\text{K}$, while the temperatures in the envelope are $T_{\text{g}} = 33.5~\text{K}$ and $T_{\text{d}} = 17~\text{K}$. The gas and dust temperatures in the central
regions of the cloud are approximately the same, since the gas is mainly cooled via radiation by dust (right panel in Fig. \ref{fig02}). In the second model, a central, cool (with a mean gas temperature of about $7~\text{K}$)
cloud core with a radius of about $0.05~\text{pc}$ and a mass of $1~M_{\odot}$ and an extended thermal envelope with a characteristic gas temperature of $12~\text{K}$ are clearly distinguished.

The cloud masses and radii are the same in the two models. This means that these models can be taken to be different stages in the quasi-static evolution of a single cloud. The equilibrium configurations obtained can be used to analyze the structures of observed starless cores, since they are more accurate than the widely used isothermal Bonnor-Ebert models.

\subsection{Collapsing protostellar clouds}

As an example, we describe the results of our numerical modeling of the compression of a protostellar cloud with initial parameters corresponding to
the first equilibrium configuration from the previous section (central density 
$n_{\text{c}} = 10^4$~cm$^{-3}$, envelope density $n_{\text{ext}} =
10^{3}$~cm$^{-3}$, radius $R_{\text{core}} = 0.2~\text{pc}$, mass $M_{\text{core}} = 4.4~M_{\odot}$). We specified in addition a uniform magnetic field ($B_0 = 2.6$~$\mu$G, ratio of magnetic energy to the modulus of the gravitational energy $0.05$) and rigid rotation ($\Omega_0 =
3.3\cdot 10^{-14}~\text{s}^{-1}$, ratio of rotational energy to the modulus of the gravitational energy $0.1$) in the cloud. This uniform magnetic field is force-free, and so does not influence the equilibrium structure. The rotation creates a centrifugal force, which perturbs the initial equilibrium state. However, we specified fairly slow rotation of the cloud, which did not lead to appreciable deviations from the equilibrium state. As the computational results showed, this demonstration model has rich physical properties that are clearly manifest in the molecular line profiles.

\begin{figure}[t]
\centering
\includegraphics[width=0.39\textwidth,angle=270]{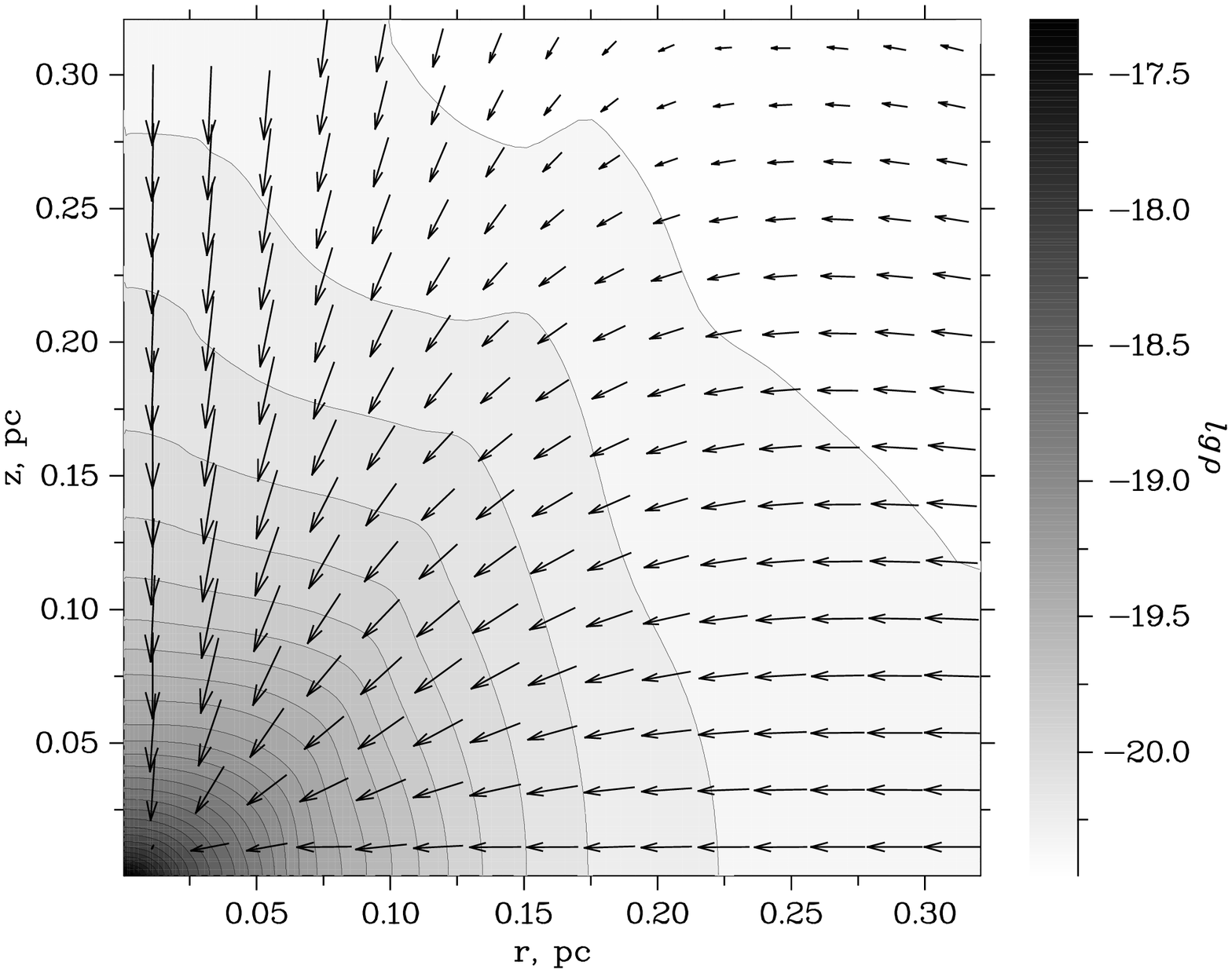}\hfill
\includegraphics[width=0.39\textwidth,angle=270]{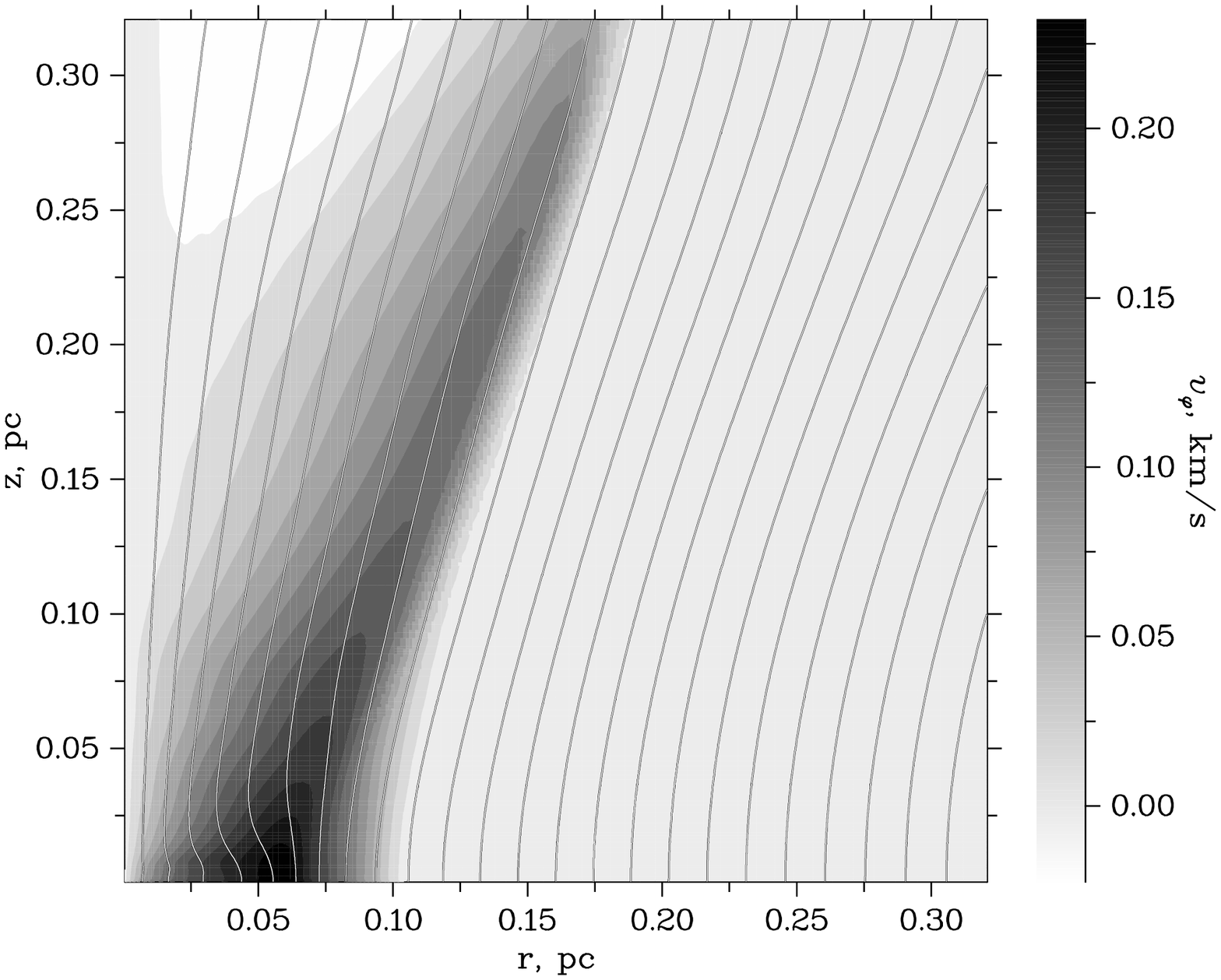}
\caption{Modeling results for the collapse of a protostellar cloud. The left panel shows the distributions of the logarithm of the density (gray scale) and the poloidal velocity (arrows), while the right panel shows the distribution of the rotational velocity (gray scale) and the magnetic field lines.} \label{map00}
\end{figure}

Figure \ref{map00} shows the distributions of the logarithm of the density (gray scale and contours) and the poloidal velocity (arrows) in the cloud at time $t = 5.8~t_{\ff} = 2$ million years from the initial contraction, where  $t_{\ff} = 1/\sqrt{4\pi G\rho_0}$ is the free-fall time and, $\rho_0$ the initial density at the cloud center. By this time, the central density has reached $n_{\text{c}} = 10^6~\text{cm}^{-3}$. The velocity of the collapse is slowed near the equatorial plane by the action of the electromagnetic and centrifugal forces. Therefore, the cloud takes on an elongated structure with time.

\begin{figure}[t]
\centering
\includegraphics[width=0.49\textwidth]{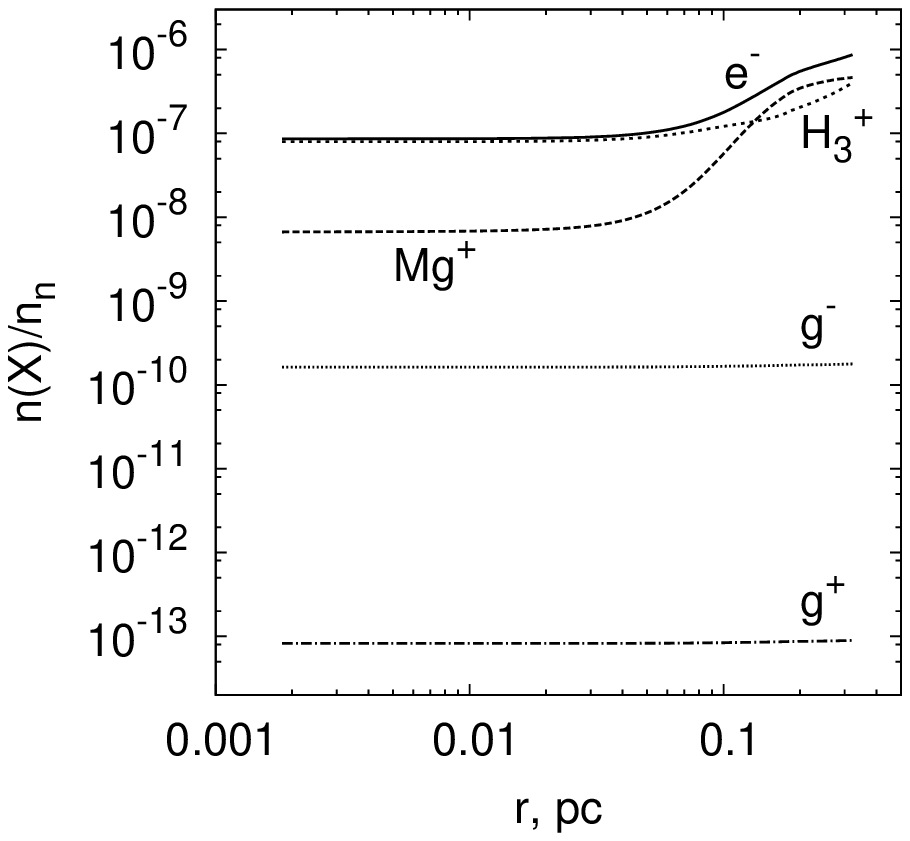}
\includegraphics[width=0.49\textwidth]{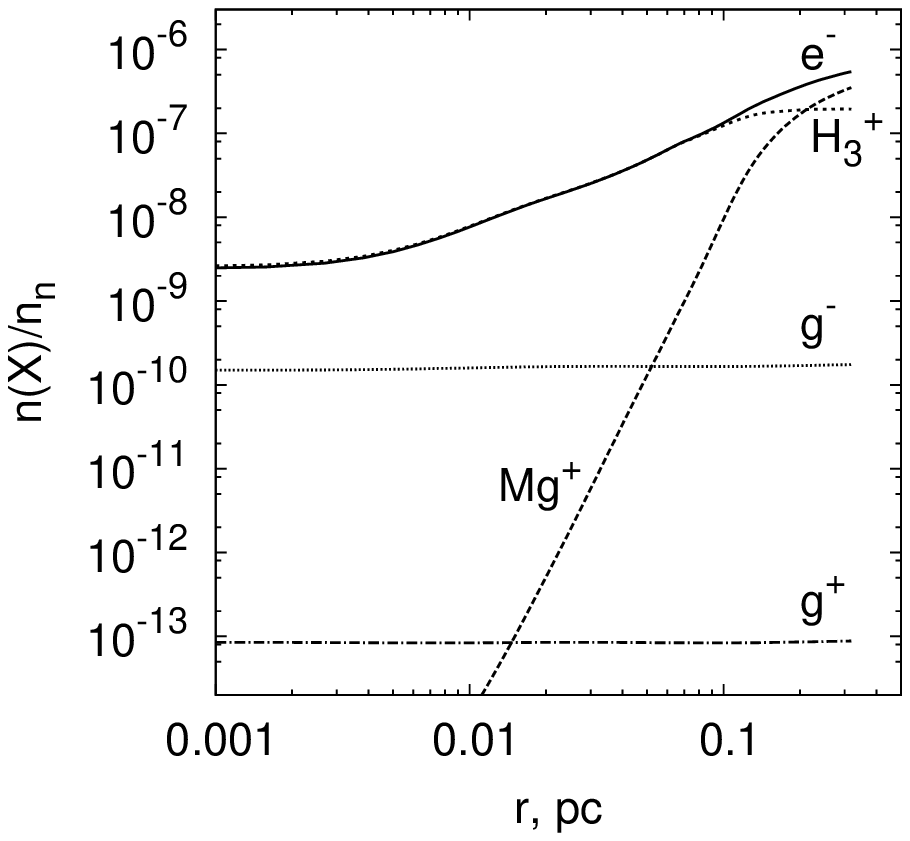}
\caption{Distributions of the abundances of charged particles in the equatorial plane for a model of a collapsing cloud for $t=0$, $n_{\text{c}} = 10^4~\text{cm}^{-3}$ (left) and $t = 5.8~t_{\ff}$, $n_{\text{c}} = 10^6~\text{cm}^{-3}$ (right)}%
\label{figIon}
\end{figure}

Figure \ref{figIon} shows distributions of the abundances of charged particles in the equatorial plane at $t=0$ and $t = 5.8~t_{\ff}$. In both cases, 
Mg$^+$ is the dominant ion in the outer layers of the cloud, where the photoionization of Mg atoms is substantial and the adsorption of Mg atoms onto dust grains is weak. In the dense central layers of the cloud, adsorption of Mg occurs on a time scale comparable to or shorter than $t_{\ff}$, and the abundance of Mg$^+$ becomes negligibly small by $t = 5.8~t_{\ff}$. The abundance of H$_3^+$ in our model is appreciably higher in all layers than in more complex models, such as that of \cite{Shematovich:2003}, providing a good agreement with the electron abundance. The abundances of charged dust grains remain nearly constant in the collapse stages considered, and are appreciably lower than the abundances of ions and electrons.

Magnetic ambipolar diffusion leads to appreciable damping of MHD waves, but only weakly influences the evolution of the large-scale magnetic field.
Therefore, a toroidal magnetic-field component, $B_{\varphi}$ is generated in the collapsing cloud due to its differential rotation. This gives rise to a braking torque that brings about a redistribution of angular momentum between the central and outer regions of the cloud. The right panel in Fig. \ref{map00} shows the distribution of the azimuthal velocity $v_{\varphi}$ (gray scale) and the magnetic field lines at time $5.8~t_{\ff}$. This shows that the region of the redistribution of angular momentum is determined by the magnetic tube formed by the toroidal magnetic field. The local specific angular momentum is redistributed within the cloud, among its central and peripheral regions, and is also carried into the external medium along the  formed magnetic tube. The opening angles of the magnetic tube and the cone of the maximum azimuthal velocity nearly coincide, and are approximately $40^{\circ}$, in agreement with theoretical estimates and observed opening angles for magnetic field lines in starless and protostellar cores \cite{Dudorov2008}.

\begin{figure}[t]
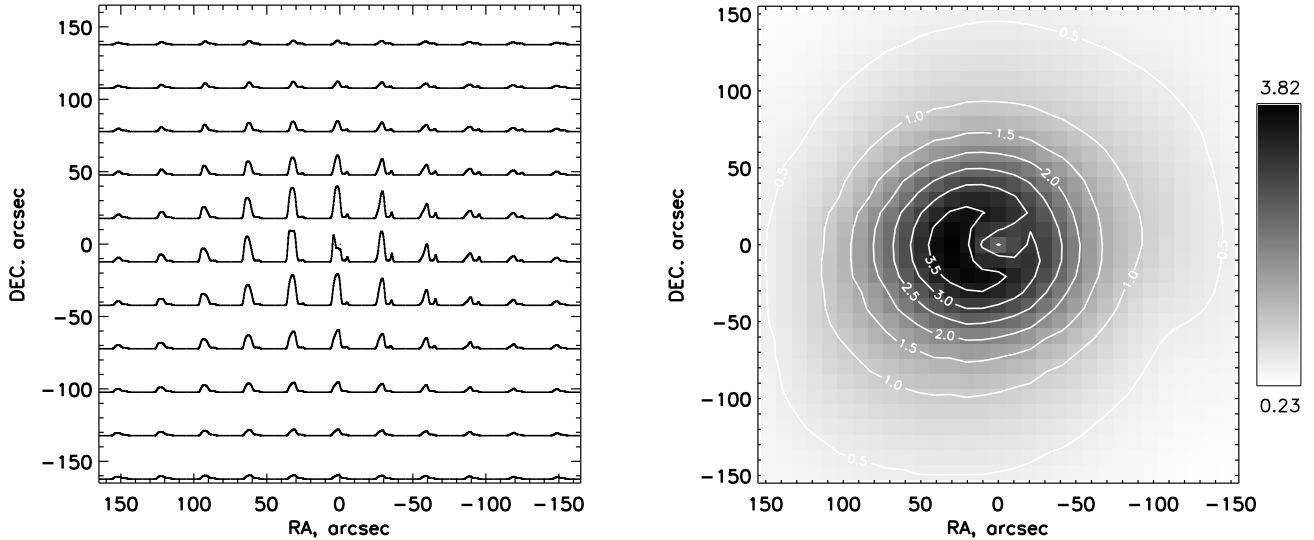

\centering
\includegraphics[width=0.445\textwidth]{fig4c.epsi}\hfill
\includegraphics[width=0.49\textwidth]{fig4b.epsi}
\caption{Results of modeling the radiative transfer in the HCO$^+$(1-0) line for a collapsing protostellar cloud. The left panel shows a map of the line profiles for the central regions of the cloud, while the right panel shows a map of the integrated radiation intensity in K km/s. The distance to the cloud is 140~pc, the inclination of the symmetry axis relative to the observer $i$=60$^{\circ}$, and the rotation of the projected axis in the plane of the sky PA=0$^{\circ}$. The asymmetry of the line profiles is due to the collapse velocity and the differential rotation of the cloud.}
\label{map01}
\end{figure}

\begin{figure}[t]
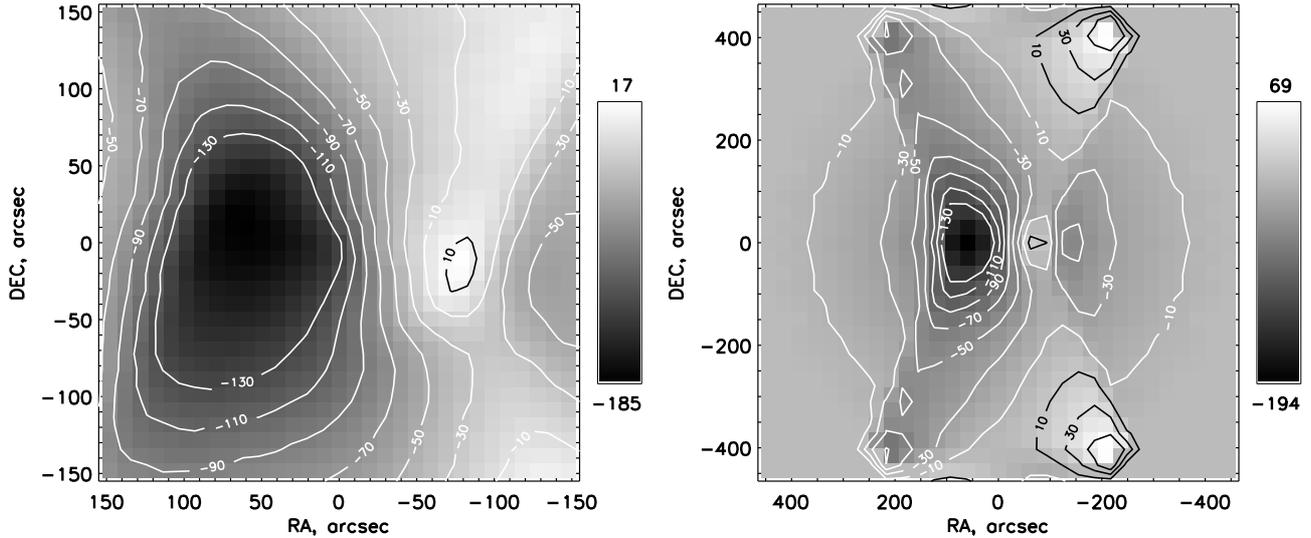

\centering
\includegraphics[width=0.49\textwidth]{fig5a.epsi}\hfill
\includegraphics[width=0.49\textwidth]{fig5b.epsi}
\caption{Results of modeling the radiative transfer in the HCO$^+$(1-0) line for a collapsing protostellar cloud. The left panel shows a map of the velocity centroid for the central regions of the cloud (the inclination angle is $i$=60$^{\circ}$). while the right panel shows a map of the velocity centroid for the entire cloud ($i$=90$^{\circ}$). The velocity is given in m/s.} \label{map02}
\end{figure}

Figure~\ref{map01} presents the results of modeling the radiative transfer in the HCO$^+$(1-0) line, which is often considered in observations of starless cores. We assume that the number density of HCO$^+$ relative to hydrogen is constant throughout the cloud and equal to $10^{-9}$, which corresponds to the typical mean values in starless cores \cite{Pavlyuchenkov2008}. We consider the case when the inclination of the symmetry axis of the cloud relative to the observer is $i$=60$^{\circ}$, and the rotation of the projected axis in the plane of the sky is PA=0$^{\circ}$. The distance to the cloud is taken to be 140~pc. We have not convolved the results with any telescope beam. We can see on the spectral map (left) that the line profiles have a complex, asymmetric shape with this asymmetry changing with position on the plane of the sky. The formation of these asymmetric profiles is due to the self-absorption of radiation in an optically thick medium with a velocity gradient along the line of sight~\cite{Pavlyuchenkov2008}. Profiles with double peaks dominate in the right-hand part of the map, with the left peak stronger than the right peak. The right peak is barely visible in the line profiles in the left-hand part of the map. This can be understood if the projections of the infall and rotation velocities along the line of sight
have different signs in the right-hand part of the map, leading to the formation of more symmetrical, two peaked profiles, while these two projected velocities have the same sign in the left-hand part of the map, so that the right peaks are essentially fully suppressed. The map of the integrated intensity (right) is also asymmetric. In particular, the position of the intensity maximum does not coincide with the direction toward the center of the cloud, due to the described variation in the line asymmetry.

Figure~\ref{map02} presents maps of the velocity centroid (the first moment of the line profile or mean velocity) for the central regions of the cloud for $i$=60$^{\circ}$ (left) and for the entire cloud for $i$=90$^{\circ}$ (right). A gradient in the mean velocity due to the rotation of the central regions of the protostellar cloud is clearly visible in the left panel. At the same time, the mean velocity is predominantly negative, testifying to a collapse, in accordance with the velocity distributions in the model. The rotational-velocity field has an hourglass shape (right panel in Fig.~\ref{map00}) and is clearly manifest in the distribution of the velocity centroid for the entire cloud (right panel in Fig.~\ref{map02}). At large distances from the cloud center, where the manifestations of the collapse are weaker, the distribution of the velocity centroid becomes more symmetrical
about the projection of the rotational axis (white and black contours). Like the opening angle for the cone representing the maximum azimuthal velocity in the dynamical model (right panel in Fig.~\ref{map00}) the opening angle for the velocity contours far from the cloud center is roughly $40^{\circ}$. The described observational manifestations of the rotation of the cloud can be used
when planning observations of starless cores. If these properties are detected, this will provide evidence for the mechanism for carrying away angular momentum described above.

\section{Conclusion}

We have presented a physical model and corresponding program package for the computation of the evolution of protostellar clouds and their observational properties. The physical model is based on magneto-gas-dynamical equations and methods for computing the thermal and ionization structure of a cloud. Ohmic and ambipolar diffusion are taken into account, as well as radiative transfer through dust. The calculated dynamical and thermal structures of clouds are used to model the radiative transfer in molecular lines. We have presented the results of modeling the structure of a quasi-equilibrium protostellar cloud and its subsequent evolution as an illustrative example.

We have not considered the chemical structure of starless cores, which could substantially influence the formation of the molecular-line profiles. Analysis
of observations of starless cores indicates that many molecules are frozen onto the surfaces of dust grains in the central regions \cite{Tafalla2002, Tafalla2004}, while molecules are destroyed by interstellar UV radiation in the cloud envelopes \cite{Pavlyuchenkov:2006}. Therefore, it is important to use a self-consistent chemical-dynamical model for modeling the corresponding line profiles and interpreting observations. The development of a module for computing the chemical evolution is a separate task that will be the subject of future work. After the addition of this chemical module, it will be possible to use our program package not only for theoretical studies of the structure of starless cores, but also for modeling of particular observed objects.

Our program package can also be used to model more advanced stages of the collapse of protostellar clouds, right up to the formation of an opaque core (protostar). However, this requires modification of the method used to compute the radiative transfer for the case of an optically thick medium.

\section*{Acknowledgments}
This work was supported by the Russian Foundation for Basic Research (project codes 06-02-16097, 07-02-01031, 08-02-00371, 07-02-96028-URAL), the Program of State Support for Leading Scientific Schools of the Russian Federation (grant NSh-4354.2008.2), and the Basic Research Programs of the Presidium of the Russian Academy of Sciences ''The Origin, Structure, and Evolution of Stars and
Galaxies'' and ''Fundamental Problems in Informatics and Information Technology''. The authors thank A.E. Dudorov, B.M. Shustov, D.Z. Wiebe, and
Th. Henning for useful discussions.

\section*{Appendix}

\appendix

\section{Adaptive mesh} \label{appA}

Let us describe the method used to construct the adaptive, moving mesh used in the two-dimensional numerical code. The theory of such meshes is based on canonical transformations of a hyperbolic system of equations in conservative form. As applied to the system (\ref{eq207}-\ref{eq209}), we will use the following transformation of coordinates and time:
\begin{equation}\label{eqa1}
  t = \tau, \,\,
  r = r(\tau, \xi), \,\,
  z = z(\tau, \zeta),
\end{equation}
which can be rewritten in differential form
\begin{equation}\label{eqa2}
  dr = \omega_r d\tau + Q_r d\xi, \,\,
  dz = \omega_z d\tau + Q_z d\zeta,
\end{equation}
where $\omega_{r,z}$ is the velocity of the mesh and $Q_{r,z}$ are metric coefficients. The law for the transformation of the coordinates (\ref{eqa1}) is chosen in the form
\begin{equation}\label{eqa3}
  t = \tau, \,\,
  r(\tau, \xi) =
  R\fr{q_r(\tau)^{\xi/\Delta\xi} - 1}{q_r(\tau)^{R/\Delta\xi} - 1}, \,\,
  z(\tau, \zeta) =
  R\fr{q_z(\tau)^{\zeta/\Delta\zeta} - 1}{q_z(\tau)^{R/\Delta\zeta} - 1}.
\end{equation}
Using these relations, we can obtain explicit expressions for the velocities of the mesh $\omega_r$ and $\omega_z$ and the metric coefficients $Q_r$ and $Q_z$.

The positions of the nodes $r_{i}(\tau)$, $z_{j}(\tau)$ depend on time, so that, generally speaking, the mesh in the initial coordinates ($r$, $z$) at an arbitrary time is inhomogeneous. It is not difficult to convince oneself that the transformation law (\ref{eqa3}) corresponds to the conditions
\begin{equation}\label{eqa4}
  \fr{\Delta r_{i+1/2}}{\Delta r_{i-1/2}} = q_r(\tau), \,\,
  \fr{\Delta z_{j+1/2}}{\Delta z_{j-1/2}} = q_z(\tau),
\end{equation}
where $\Delta r_{i+1/2} = r_{i + 1} - r_{i}$ and $\Delta z_{j+1/2} = z_{j + 1} - z_{j}$ are the mesh steps in the $r$ and $z$ directions. Thus, the
quantities $q_{r,z}(\tau)\ge 1$ are parameters of the mesh compression in the corresponding directions. When $q_{r,z} = 1$, the mesh in ($r$, $z$) is uniform, since then $r_i = \xi_i$, $z_j = \zeta_j$.

The time dependence of the compression parameter can be specified arbitrarily taking into account the assumed character of the flow. The contraction of protostellar clouds has a characteristic dynamical time scale --- the free-fall time $t_{\ff} = 1/\sqrt{4\pi G\rho_0}$, where $\rho_0$ is the initial density at the cloud center. We used the following expression for the compression parameter:
\begin{equation}\label{eqa5}
  q(\tau) = q_{\infty} -
  \left(q_{\infty} - q_0\right)
  \left(1 + \tau/\tau_{*}\right)
  e^{-\tau/\tau_{*}},
\end{equation}
where $q_0$, $q_{\infty}$, and $\tau_{*}$ are parameters. This dependence
was chosen so that $q(0) = q_0$, $q \to q_{\infty}$ as $\tau \to \infty$, and variations of the function $q(\tau)$ occur on the characteristic time scale $\tau_{*}$. Moreover, $\dot{q}(0) = 0$. The initial and final values of the compression parameters $q_r(\tau)$ and $q_z(\tau)$ should not be very different. We chose these so that the relative difference in the sizes of neighboring mesh cells in ($r$, $z$) did not exceed 10-20\%. The characteristic time scale $\tau_{*}$ was chosen to be several free-fall times $t_{\ff}$.

\small%

\normalsize

\end{document}